# MotorBeat: Acoustic Communication for Home Appliances via Variable Pulse Width Modulation

WEIGUO WANG, Tsinghua University, China
JINMING LI, Tsinghua University, China
YUAN HE*, Tsinghua University, China
XIUZHEN GUO, Tsinghua University, China
YUNHAO LIU, Tsinghua University, China

More and more home appliances are now connected to the Internet, thus enabling various smart home applications. However, a critical problem that may impede the further development of smart home is overlooked: Small appliances account for the majority of home appliances, but they receive little attention and most of them are cut off from the Internet. To fill this gap, we propose MotorBeat, an acoustic communication approach that connects small appliances to a smart speaker. Our key idea is to exploit direct current (DC) motors, which are common components of small appliances, to transmit acoustic messages. We design a novel scheme named Variable Pulse Width Modulation (V-PWM) to drive DC motors. MotorBeat achieves the following 3C goals: (1) *Comfortable* to hear, (2) *Compatible* with multiple motor modes, and (3) *Concurrent* transmission. We implement MotorBeat with commercial devices and evaluate its performance on three small appliances and ten DC motors. The results show that the communication range can be up to 10 m.

## 1 INTRODUCTION

The recent years have witnessed rapid advances in smart home [18, 60]. Many home appliances are now connected to the Internet and thus are endowed with interesting capabilities to improve the user experience. The users can interact with them remotely, and these appliances themselves can cooperate in providing a seamless service [30]. Despite the trend of smart home, there are still a considerable number of home appliances cut off from the Internet, especially small appliances. Home appliances can be mainly classified into three categories [72]: small appliances, major appliances (i.e., white goods), and consumer electronics (i.e., brown goods). In contrast to major

*Corresponding author

Authors' addresses: Weiguo Wang, wwg18@mails.tsinghua.edu.cn, Tsinghua University, Beijing, China; Jinming Li, li-jm19@mails.tsinghua.edu.cn, Tsinghua University, Beijing, China; Yuan He, heyuan@tsinghua.edu.cn, Tsinghua University, Beijing, China; Xiuzhen Guo, guoxiuzhen94@gmail.com, Tsinghua University, Beijing, China; Yunhao Liu, yunhao@mail.tsinghua.edu.cn, Tsinghua University, Beijing, China.

Table 1. The sales, prices, ratios, and status indicators of small appliances.

| Group | Keyword | Sales[1] (k) | Price ($) | Ratio[2] | Status Indicators[3] |
|---|---|---|---|---|---|
| [illegible] | Electric Toothbrush | 384.4 | 66.9 | 3% | charge-battery, replace-head [45] |
| | Hair Trimmer | 280.4 | 16.4 | 0% | charge-battery, clean-head, replace-head, add-oil [36] |
| | Electric Razor | 194.4 | 60.9 | 0% | charge-battery, clean-head, replace-head [47] |
| | Hair Dryer | 182.6 | 46.5 | 0% | mode |
| [illegible] | Rice Cooker | 632.0 | 120.9 | 0% | job-done, working-mode |
| | Electric Mixer | 151.5 | 36.1 | 0% | replace-blade |
| | Coffee Maker | 135.6 | 94.4 | 0% | job-done, working-mode |
| | Blender | 116.5 | 210.6 | 0% | replace-blade [49] |
| [illegible] | Humidifiers | 366.9 | 45.2 | 3% | clean-filter, replace-filter, add-water [46] |
| | Massager | 340.2 | 90.8 | 1% | charge-battery, working-mode |
| | Vacuum Cleaner | 278.1 | 196.5 | 5% | clean-filter, replace-filter, clean-dust-bag [50] |
| | Fan | 198.0 | 68.9 | 2% | charge-battery, working-mode |
| | Average | 271.8 | 87.8 | 1.2% | - |

[1]The sales are calculated as the total number of customer reviews.
[2]The ratio denotes the percentage of small appliances that can transmit data.
[3]The status indicators are collected from user manuals, technical specifications, or LED lights of appliances' appearances.

appliances such as refrigerators and washing machines, small appliances mainly refer to portable or semi-portable household devices. Examples include electric toothbrushes, blood pressure monitors, and fans.

One basic fact is that small appliances now account for a major proportion of home appliances, and this proportion is expected to continue to grow. According to statistics in 2020 [62], the sales volume of small appliances is 8.3x larger than that of major appliances in the USA. Besides, small appliances are dedicated to meeting consumers' various and fine-grained demands, while major appliances mainly focus on accomplishing necessary housework tasks. This means that the demand for small appliances increases with the quality of life.

However, the vast majority of small appliances are not connected to the Internet. We investigate three groups of small appliances, shown in Table 1. We separately select the four most popular appliances in each group based on their sales. The names of these appliances are used as keywords to search for the top 100 best-selling items on Amazon [3]. Table 1 lists the ratio of small appliances that are connected to the Internet. As we can see, most small appliances are isolated from the Internet. The average connection ratio is only 1.2%.

In summary, an overlooked but critical problem is that small appliances account for the majority of home appliances, but they receive little attention, and few of them are connected to the Internet.

In this paper, we present a new approach MotorBeat that connects small appliances to the smart speaker. As illustrated in Figure 1, the idea is to exploit electric motors, which are widely installed on small appliances, to talk to the smart speaker in the acoustic channel. The idea of MotorBeat stems from two key observations:

The first observation is that nearly 80% of small appliances contain an electric motor[1]. This observation ensures that MotorBeat can be widely applied to most small appliances without significant hardware modification. The second observation is that smart speakers are now very popular, which have access to the Internet and contain always-on microphones. Smart speakers can then behave as free gateways for appliances by receiving and relaying their acoustic messages to the Internet.

The ability of small appliances to talk to the smart speaker will give birth to many applications. (1) Small appliances can report their statuses, such as ON-OFF state, remaining battery, and working mode. Table 1 lists the detailed status indicators that small appliances may report. Based on the ON-OFF state of, for example, your electric toothbrush, smart home systems can know when and how long you brush your teeth, and remind you when you forget to do so. Similarly, smart home systems can also warn you to turn off some appliances, like hair dryers, with abnormal running time to avoid fire [26]. (2) Such ability may also be attractive to appliance manufacturers. If the status of their products can be continuously acquired, they can track the entire life cycle of each appliance [12, 51], based on which they can provide better after-sale service and improve the future product designs. (3) From a general perspective, smart home systems can explicitly or implicitly perceive more contextual information of home environments [2, 40], such as human activity, appliance usage, and further our daily life.

[1]We check all 56 default small appliances presented on Amazon [4], and 44 of them contain an electric motor (i.e., 44/56 $\approx$ 78.6%).

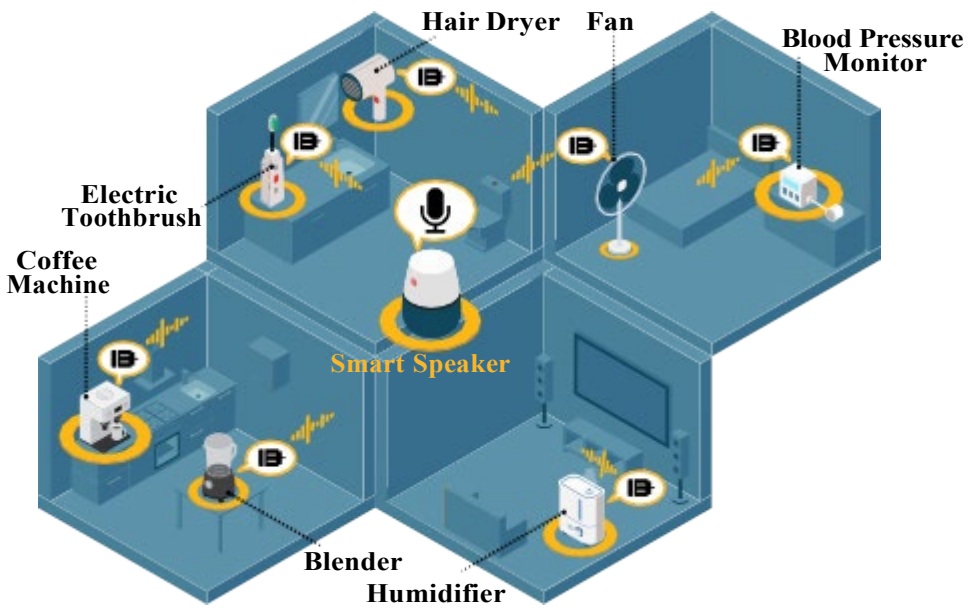

Fig. 1. An illustration of MotorBeat.

We notice that a recent pioneering work, Bleep [8], also exploits motors to enable UAVs to communicate with each other. The solution of Bleep is acceptable in industrial environments, but is unreasonable in home environments: First, Bleep encodes information in linear chirps and generates sounds that are unfriendly to human ears. Second, Bleep equips each UAV with an additional microphone to sense channels and choose an idle channel. In our case, small appliances hardly contain a microphone to avoid collisions, and it's unrealistic to additionally equip each appliance with a microphone. In our home environments, MotorBeat should achieve the following **3C** (Comfortable, Compatible, and Concurrent) goals:

- **Comfortable** to hear. Small appliances are designed to serve customers. The quality of experience is critical for them. Therefore, MotorBeat should not disturb the users while transmitting acoustic messages.
- **Compatible** with multiple motor modes. The motors of small appliances typically have multiple working modes, and these modes could be changed unpredictably by the users at any time. This requires that the transmission of MotorBeat should not affect the function of these motors, and the transmission itself should not be disturbed by users.
- **Concurrent** transmission. Small appliances typically have no microphone to monitor the transmission of other appliances. This means that collision detection or collision avoidance is infeasible for small appliances, and collisions are inevitable in our scenario. To tolerate the collisions, MotorBeat should allow concurrent transmission and support decoding from collided signals.

To achieve the above **3C** goals, MotorBeat introduces a novel modulation technique, Variable Pulse Width Modulation (V-PWM). At a high level, small appliances drive the voltage of their motors with a pseudo-random (i.e., variable) switching frequency. By doing so, the harmonics of the acoustic signals, which sound uncomfortable, are significantly reduced (**Comfortable**). Furthermore, even though the switching frequency is variable, the duty cycle of the voltage can be configured online without impeding the transmission, thus supporting multiple working modes of motors (**Compatible**). Last, we assign each appliance with a unique V-PWM symbol. These symbols are orthogonal to each other. Similar to Code-Division Multiple Access (CDMA), multiple appliances are allowed to transmit concurrently. The MotorBeat receiver is able to separately detect and decode messages of these appliances (**Concurrent**). Our contributions are as follows:

- We propose MotorBeat, a novel motor-based communication paradigm that enables small appliances to talk to a smart speaker. By doing so, we can connect small appliances to the Internet.
- We disclose the acoustic characteristics of DC motors, and show the opportunity to communicate based on them. We introduce a novel modulation technique, V-PWM, to drive the motors and achieve **3C** goals required in the real world.
- We implement MotorBeat and evaluate its performance on three small appliances (electric toothbrush, blood pressure monitor, and fan) and ten different DC motors. The results show that MotorBeat can be widely applied to small appliances. The communication range can be up to 10 m, ensuring that MotorBeat can cover a standard apartment.

**Roadmap.** Sections 3 and 4 introduce how a DC motor works and its acoustic characteristics. Section 5 gives MotorBeat's overview. In Sections 6 and 7, we elaborate on the design of transmitter and receiver, respectively. Section 8 discusses some practical issues. Section 9 presents the implementation and evaluation results. Section 2 discusses the related work. Sections 10 and 11 respectively discuss and conclude this work.

## 2 RELATED WORK

**Motor-Based Communication.** Bleep [8] is the work closest to ours. Bleep modulates the sounds of UAV motors to enable UAVs to communicate in the acoustic channel. Bleep increases or decreases the switching frequency of PWM voltage every 50ms to transmit acoustic up-chirp or down-chirp signals. To capture and decode the acoustic signals, Bleep equips each UAV with an additional microphone.

Ripple [54, 55] exploits Linear Resonant Actuators (LRAs) in smartphones to achieve motor-accelerometer communication. Similarly, VibroComm [73] utilizes LRAs to transmit vibration messages to gyroscopic sensors, thus achieving targeted and explicit communication. Differing from the existing works that use AC motors, MotorBeat's design is based on DC motors. The existing works allow to modulate both the magnitude and (or) frequency of vibrations, and have much more space for modulation. In comparison, MotorBeat satisfies more constraints in modulation, so as to preserve the original function of small appliances.

**Side-Channel of Vibration.** Acoustic signals in the side-channel of vibration have received research attention in recent years. For example, ViBand [34] and SecureVibe [32] explore bio-acoustics sensed by wearable devices like smartwatch for interaction and secure communication, respectively. Deaf-Aid [19] utilizes ultrasonic signals to make the gyroscope resonate and then to convey information. Those side-channels can also be used for device authentication [14, 37, 42, 71, 75], object identification [24], drone identification [52], near-field communication [43], earshot communication [39] and inter-vehicular communication [53]. Some existing works demonstrate that the side-channels can be used to recover information from input/output devices such as keyboards [6, 25, 38], printers [7], and screens [21, 29]. In this paper, we disclose the acoustic characteristics of DC motors, and demonstrate the feasibility of using motors to transmit acoustic messages by modulating the voltage.

**Timing-Based Encoding.** As we will see soon, MotorBeat can also upload bits by embedding bits into intervals between V-PWM symbols. This method is inspired by the following works: ONPC [41] uses the timing of 802.11 frames to convey information. WiChronos [56] encodes information in the time interval between two narrow-band symbols to achieve low-power communication. FreeBee [31] achieves cross-technology communication via embedding bits into beacons by shifting transmission timing. Encoding information within the beacon timing is also studied in optical and UWB communications[16, 20, 22].

**Random Pulse Width Modulation.** Our V-PWM is inspired by random pulse width modulation which is originally designed for voltage-controlled power electronic converters [9, 35, 64]. Because this modulation can disperse the concentrated energy of harmonics over a wide frequency range, many works utilize it to mitigate acoustic noise [10, 33], electromagnetic interference [63, 65], or mechanical resonance frequency [11].

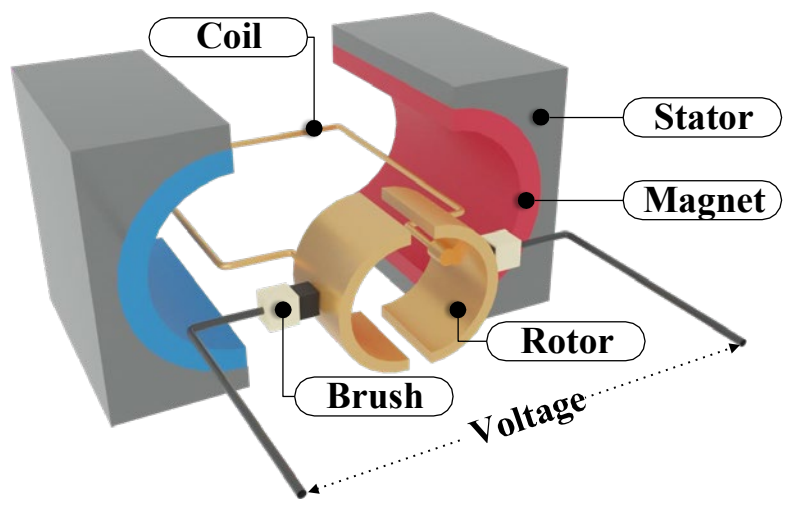

Fig. 2. DC motor structure.

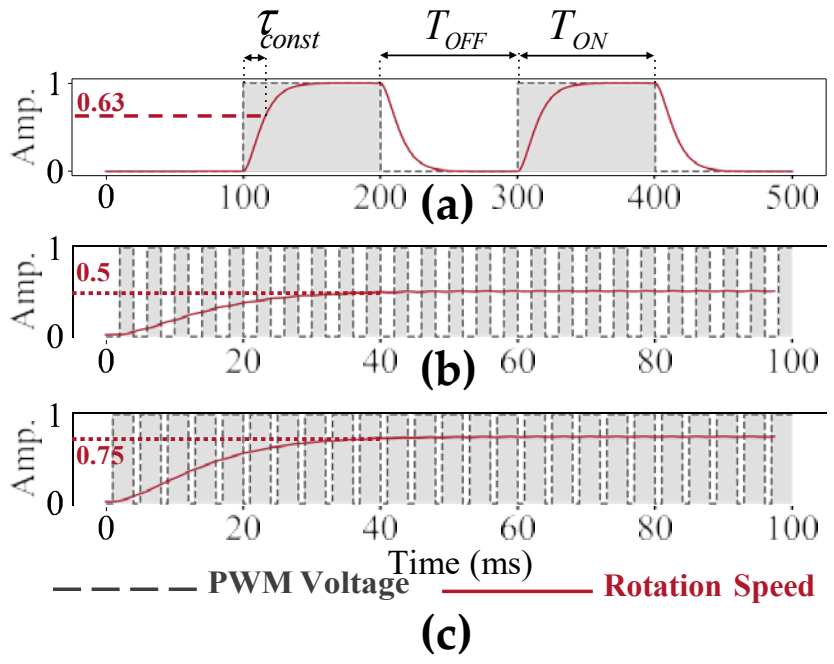

Fig. 3. PWM voltage vs. speed.

# 3 PRIMER

## 3.1 DC Motor

A direct current (DC) motor[2] shown in Figure 2 consists of two main components: stationary part (stator) and rotating part (rotor). The stator contains permanent magnets on the side, and the rotor comprises coils. The working process of a DC motor is as follows. According to the Lorentz law, when the DC motor is powered, an additional magnetic field will be created inside the rotor. The rotor is then attracted or repelled by the magnets on the stator. Two electronic brushes connect the rotor to the voltage. These brushes can switch the direction of the coil current during rotation and thus switch the polarity of the electromagnet. By doing so, the rotor can keep rotating in the same direction.

## 3.2 PWM Voltage

Pulse Width Modulation (PWM) voltage is widely adopted to drive small DC motors for its simplicity. PWM voltage is a periodic signal with two states, ON and OFF, and is described by two parameters: (1) *duty cycle* $\alpha$ and (2) *switching period* $T_{sw}$. Here, duty cycle $\alpha = \frac{T_{ON}}{T_{ON} + T_{OFF}}\%$, where $T_{ON}$ and $T_{OFF}$ denote the durations of states ON and OFF within a period, respectively (see Figure 3(a)). Another parameter switching period $T_{sw}$ is defined as $T_{ON} + T_{OFF}$. The DC motor can rotate steadily when the switching frequency $f_{sw}$ satisfies the following condition:

$$f_{sw} = \frac{1}{T_{sw}} \gg \frac{1}{\tau_{const}} \quad (1)$$

where $\tau_{const}$ denotes the **time constant**[3] of a DC motor and it can be used to quantify how fast a DC motor responds to the input voltage (see Figure 3(a)).

One way to interpret Equation 1 is to model the DC motor as a circuit with a resistor and an inductor in series, which is equivalent to a first-order low pass filter with cut-off frequency $1/\tau_{const}$. Therefore, as long as the input voltage frequency $f_{sw}$ is larger than the cut-off frequency $1/\tau_{const}$, the DC motor will rotate steadily.

The advantage of PWM is that by merely changing the duty cycle $\alpha$, the appliance can easily control its motor's rotation speed. In Figure 3(b), the duty cycle is 50%, and the switching frequency is 50x larger than that in Figure 3(a). After a short transient period (0-40ms), the motor exhibits a steady speed, 50% of its maximum speed. In Figure 3(c), the duty cycle increases to 75%, and the steady speed increases to 75% correspondingly[4].

[2]Here, we take brushed motors as an example to introduce DC motors.
[3]Time constant is formally defined as the time the motor takes to reach its 63% ($\approx 1 - 1/e$) maximum speed.
[4]In practice, the rotation speed is not a linear function of the duty cycle.

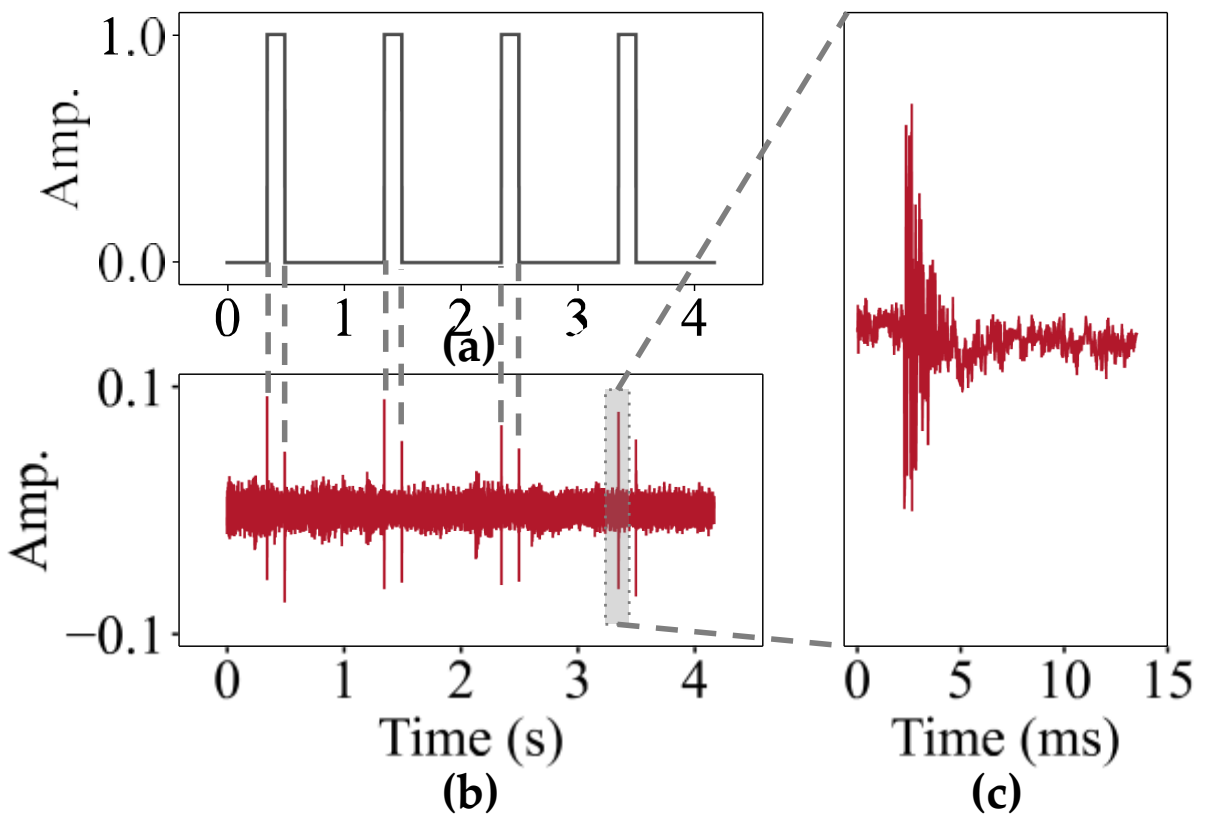

Fig. 4. (a) The voltage. (b) The acoustic signal. (c) The spike (EI signal).

# 4 SIGNAL MODEL

## 4.1 Acoustic Characteristics of DC Motors

The key idea of MotorBeat is to modulate the motor's sound in the way of modulating its voltage. In general, DC motors generate two types of acoustic signals during rotation [23, 28]:

- **Mechanism-Induced (MI) acoustic signal** is mainly caused by the rotor unbalance, and the brush friction. On the one hand, the misalignment between the rotor's inertia axis and its geometric axis will make the supporting structure vibrate and generate sound. On the other hand, the friction between the brushes and the rotor also makes a sound.
- **Electromagnetism-Induced (EI) acoustic signal** is mainly caused by the change of magnetic fields. The magnets on the stator generate a static magnetic field, and the coils on the rotor generate a dynamic magnetic field, controlled by the voltage. These two magnetic fields will make the stator and the rotor attract or repel each other, further causing the slight deformation of the motor and the vibration.

## 4.2 Identification of Signal and Noise

We select the EI signal to transmit messages, while the MI signal is the noise to our system. The considerations of our selection are following:

- Given the constraint that the motor's rotation should not be affected (**Compatible**), we can only modulate the EI signal, rather than the MI signal. Compared with the MI signal, which is highly determined by the rotation, the EI signal is independent of the rotation. This property provides us an opportunity for modulation without disturbing the rotation.
- Even if we can modulate the MI signal, the MI signal is less predictable and controllable than the EI signal. This is because the rotation speed will change with the load (or force) onto the motor, and the load is unpredictable and dynamic (e.g., the force applied to the toothbrush). As a result, the rotation and the resulting MI signal are unpredictable. On the other hand, the EI signal is only determined by the voltage. As long as the voltage is deterministic, the EI signal is deterministic.

In MotorBeat, we modulate the voltage with a V-PWM symbol (introduced later in Section 6) to make the motor generate a specific sound. In this sound, only the EI signal $S_{EI}$ can be modulated by the voltage and thus $S_{EI}$ is our desired signal, while the MI signal $S_{MI}$ will not be modulated and thus $S_{MI}$ is a noise.

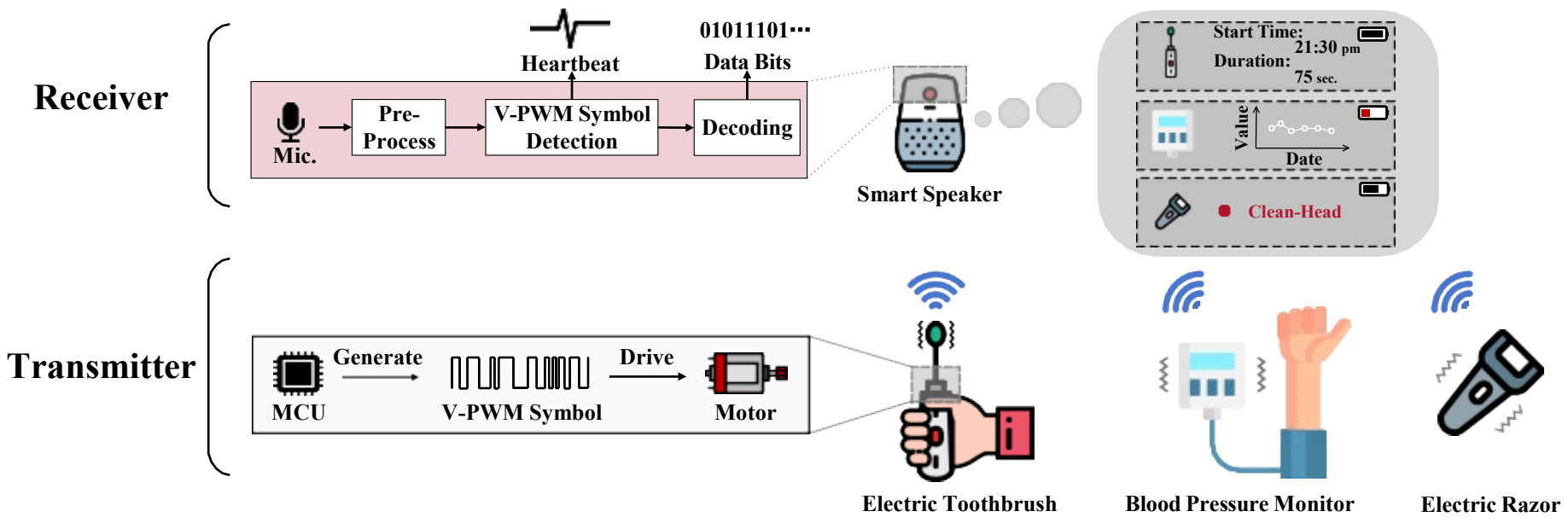

Fig. 5. Overview of MotorBeat.

To further understand the relation between the EI signal and the voltage, we especially illustrate the voltage and the corresponding acoustic signal in Figure 4(a) and (b), respectively[5]. As we can see, the edges of the voltage will introduce spikes in the acoustic signal. Let us take the positive edge (from OFF to ON) as an example to explain this: At the moment when the voltage is switched ON, the electromagnetic force will be generated and introduce a deformation of the motor. The sudden appearance of this deformation behaves like a physical knock onto the motor, thus making the sound (see Figure 4(c)). Similarly, when the voltage is switched OFF, the deformation will suddenly disappear and also behave like a physical knock.

On the other hand, the MI signal has no relation with the modulated voltage and thus is a noise to MotorBeat. The MI signal is mainly determined by the rotation. As introduced in Section 3.2, if $f_{sw} \gg 1/\tau_{const}$, the rotation speed is constant[6], which is uncorrelated with the modulated voltage.

The signal-to-noise ratio (SNR) in MotorBeats can be calculated by

$$SNR_{dB} = 10 \log \frac{P_{S_{EI}}}{P_{S_{MI}} + P_N} \tag{2}$$

where $N$ denotes the ambient noise, and $P$ denotes the signal power, specified by its subscript. In practice, the EI signal is usually buried by the noise. Firstly, for many small appliances, their DC motors are low-power devices and generate very weak sounds (both $P_{S_{EI}}$ and $P_{S_{MI}}$ are relatively small). Secondly, the sound of a DC motor is dominated by the MI signal [15], which means the EI signal is typically overwhelmed by the MI signal ($P_{S_{EI}} << P_{S_{MI}}$). From our measurements, the SNR can be lower than -15 dB when a microphone is placed only 3 cm away from a DC motor to receive the signal.

## 5 MOTORBEAT OVERVIEW

Figure 5 illustrates the overview of MotorBeat. The MotorBeat transmitters (Section 6) are various small appliances that contain DC motors. Each appliance is assigned a predefined and unique V-PWM symbol. According to its symbol, each appliance drives its motor to transmit acoustic messages. The MotorBeat receiver (Section 7) is the smart speaker, which knows the V-PWM symbols of appliances. The receiver then takes these symbols as templates to correlate with the acoustic samples to separately detect whether these symbols exist in the acoustic signal and further decode data bits. By doing so, the receiver then knows the status of home appliances, such as when and how long your electric toothbrush works, what the readings of the blood pressure monitor are, and whether your electric razor's shaving head needs to be cleaned or replaced.

[5] To clearly display the EI signal, we record the EI signal in very controlled conditions: (1) The rotor is deliberately fixed to avoid the MI signal introduced by the rotation. (2) The switching period of the voltage is enlarged to 1s to avoid the overlaps between the spikes. (3) To record the weak signal $S_{EI}$, the microphone is placed close to the motor.

[6] Even though the speed is constant, the MI sound is not a narrow-band signal, but a wide-band signal. This means that it is infeasible to separate the EI signal from the MI signal in the frequency domain.

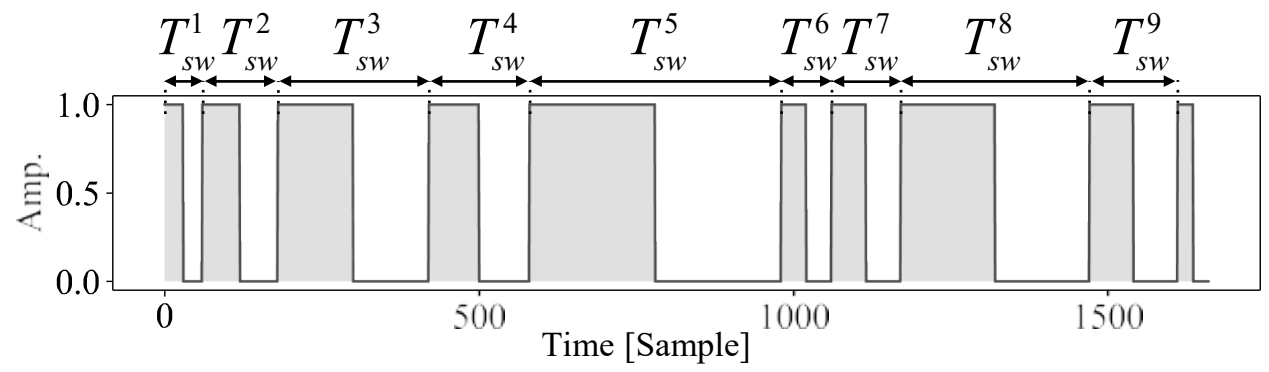

Fig. 6. Illustration of a V-PWM symbol.

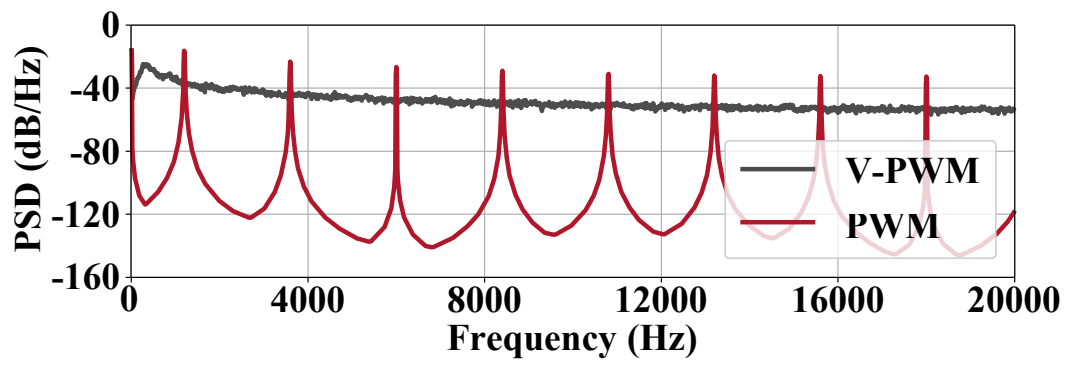

Fig. 7. Power spectral density of V-PWM and PWM.

## 6 TRANSMITTER DESIGN

In this section, we first introduce the V-PWM based modulation, and separately explain how to achieve the **3C** goals with V-PWM symbols. Then, we introduce how to transmit data bits.

### 6.1 V-PWM based Modulation

Variable PWM (V-PWM) is an extension of PWM, but randomizes the switching periods of PWM. Figure 6 shows an example of a V-PWM symbol. A V-PWM symbol is composed of a sequence of pulses. Different from PWM whose pulses are identical, the pulses of V-PWM are instead different from each other. In Figure 6, the switching period ($T^i_{sw}$) of each pulse is variable, but the duty cycle is the same (50%).

Mathematically, a V-PWM symbol is defined as a tuple $< \mathsf{A}, \mathsf{T}_{sw} >$, where $\mathsf{A} = \{\alpha^1, \alpha^2, \alpha^3, \ldots\}$ denotes the duty cycles of each pulse, and $\mathsf{T}_{sw} = \{T^1_{sw}, T^2_{sw}, T^3_{sw}, \ldots\}$ denotes the switching periods of each pulse. In the rest of this paper, we use V-PWM to refer to PWM with a fixed duty cycle but with a variable switching period. Namely, for each $\alpha^i \in \mathsf{A}$ and $T^i_{sw} \in \mathsf{T}_{sw}$, $\alpha^i$ is a constant value, while $T^i_{sw}$ is variable.

To generate a unique V-PWM symbol for each appliance, we assign each appliance with a unique ID, and use this ID as the random seed $s$ to initialize the random state to generate a sequence of pseudo-random switching periods $\{T^1_{sw}, T^2_{sw}, T^3_{sw}, \ldots\}$. Here, $T^i_{sw}$ is uniformly distributed between the minimum switching period $T^{min}_{sw}$ and the maximum switching period $T^{max}_{sw}$, i.e., $T^i_{sw} \sim U(T^{min}_{sw}, T^{max}_{sw})$.

To avoid damaging DC motors, the maximum switching period $T^{max}_{sw}$ should be smaller than the time constant $\tau_{const}$ of a DC motors. According to Equation 1, if the switching period is larger than $\tau_{const}$, DC motors cannot rotate steadily and will produce extra friction between the rotor and the stator, thus causing more wear and tear. To ensure the motor can rotate steadily, we set $T^{max}_{sw}$ to a value that is smaller than $\tau_{const}$ and therefore each $T^i_{sw}$ is smaller than $\tau_{const}$.

### 6.2 3C goals

#### 6.2.1 *Comfortable to Hear.*

Small appliances are dedicated to providing good experiences for users, so Quality of Experience (QoE) is a critical factor of small appliances. This requires that MotorBeat should not disturb the users while driving the motor, and the modulated EI signal $S_{EI}$ should be imperceptible to the users.

One solution to modulate $S_{EI}$ is Frequency-Shift Keying (FSK). Specifically, by selecting different switching frequencies of PWM voltage, the motor can emit the EI signal embedded with specific information. For example, Bleep [8] is a recent work that enables the UAV motors to communicate. To convey a symbol, Bleep increases $f_{sw}$ every 50 ms to transmit a chirp-like signal (a series of sounds with increasing but discrete frequencies). We observe that the modulated signal in Bleep will sound like "Do-Re-Mi-Fa" syllables, which is too conspicuous to neglect even though the EI signal is weak (we will evaluate this in Section 9.6). This is because, from a physiological perspective, human ears are sensitive to sounds whose energy is concentrated at discrete frequencies [44].

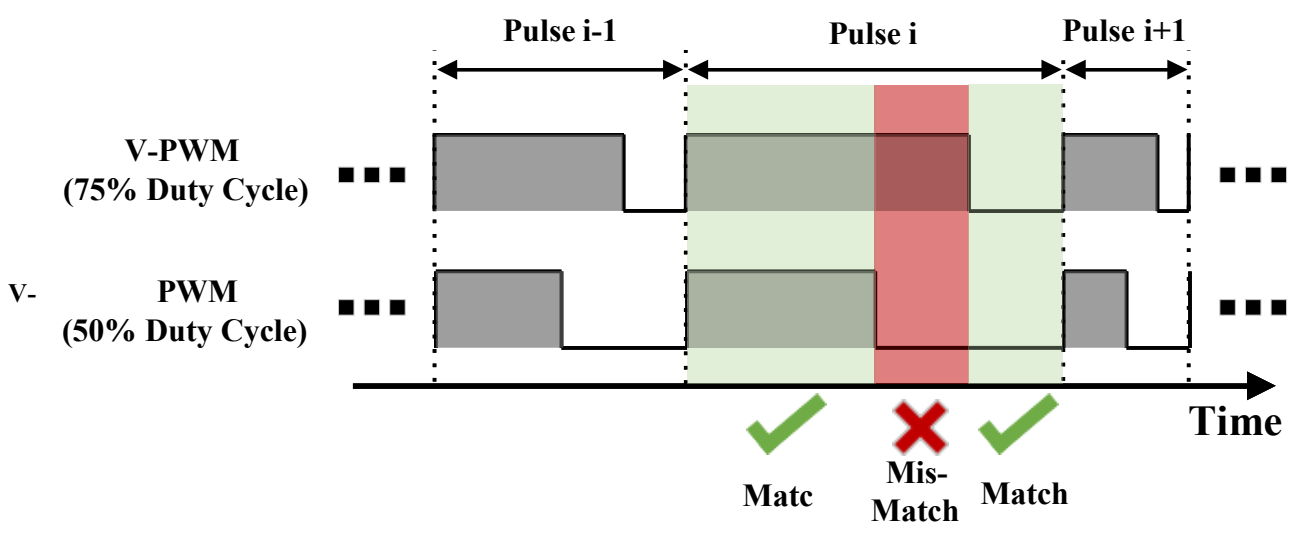

Fig. 8. As long as two V-PWM symbols have the same switching periods, they are highly correlated even if their duty cycles are different.

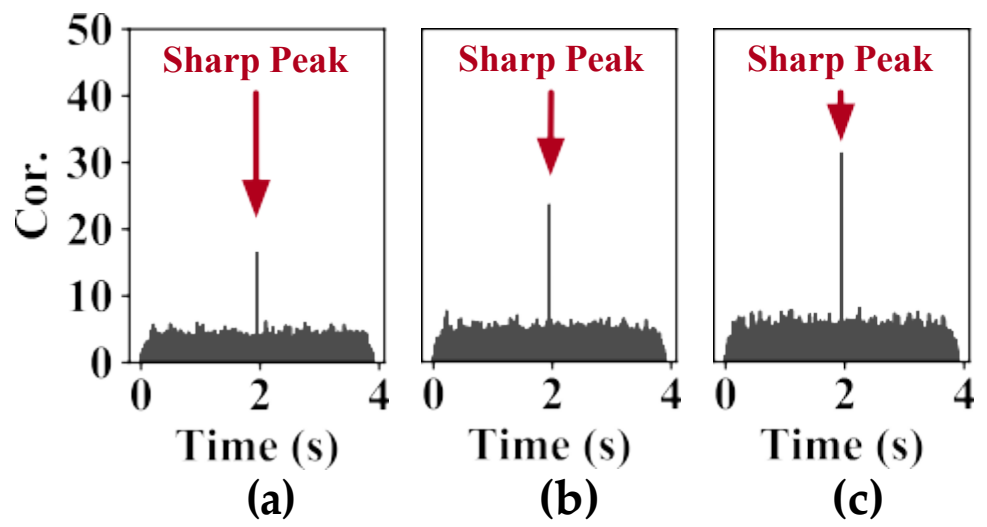

Fig. 9. Correlations between symbols with different duty cycles: Correlations between (a) $S_{20}$ and $S_{50}$, (b) $S_{30}$ and $S_{50}$, and (c) $S_{40}$ and $S_{50}$.

Figure 7 shows the Power Spectral Density (PSD) of a PWM signal. As we can see, the energy of this signal is mainly distributed in several discrete harmonics. Therefore, the sound in Bleep is annoying to the users. Of course, such sound is acceptable for UAVs designed to work in noisy industrial environments, but is unacceptable for appliances designed to work in quiet home environments.

We adopt V-PWM method to modulate the EI signal, which is imperceptible to the users in MotorBeat. The key idea is that V-PWM randomly changes each switching frequency (i.e., $1/T^i_{sw}$), and distributes the energy over a wide frequency range (see Figure 7). Therefore, the discrete harmonics of the resulting EI signal are significantly reduced. The EI signal is then converted to a white-noise-like signal, which is friendly to human ears.

---

**Algorithm 1:** Online Transmit an V-PWM symbol

---

Use appliance ID to initialize the random state, and generate a sequence of pseudo-random switching periods $\mathrm{T}_{sw}$;
Initialize duty cycle $\alpha$;
Listen for button interrupt with a handler *Callback*;
**def** *Callback (mode m)*
  Update duty cycle $\alpha$ to a new value that corresponds to mode $m$;
**end**
**foreach** $T^i_{sw}$ *in* $\mathrm{T}_{sw}$ **do**
  Set voltage ON, and sleep $\alpha \cdot T^i_{sw}$ seconds;
  Set voltage OFF, and sleep $(1-\alpha) \cdot T^i_{sw}$ seconds;
**end**

---

6.2.2 *Compatible with Multiple Modes.* Many small appliances have different working modes. For example, the electric fan may have multiple operating speeds; The electric toothbrush usually supports multiple modes, such as `sensitive` and `clean`, to provide different intensities [48]. Moreover, the users can change these modes at will by pressing the button, which means that the motor mode can change unpredictably. These facts require that MotorBeat should support multiple working modes of appliances without disturbing their basic function.

Our solution is simple. The appliances generate and transmit the V-PWM symbol on the air. Algorithm 1 demonstrates the simplified procedures. When a user pushes the button to select a new mode, the handler function `CallBack` will be triggered to update the duty cycle $\alpha$. Therefore, the motor speed will quickly change according to the user interaction. Typically, a larger duty cycle corresponds to a higher rotation speed.

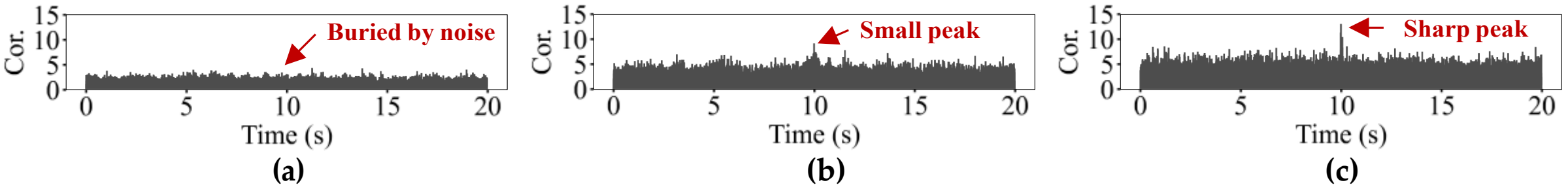

Fig. 10. Correlations with different symbol lengths: (a) 0.1s. (b) 0.5s. (c) 1s.

One may wonder, since the duty cycle has changed when an V-PWM symbol is being transmitted, will the actually transmitted symbol mismatch with the symbol previously known to the receiver, resulting in the failure of symbol detection?

The answer is no. The reason is the following. The receiver detects the transmitted V-PWM symbol of each appliance by correlating its previously-known symbol with the acoustic signal. This means that, provided that the transmitted symbol is highly correlated with this previously-known symbol, the receiver can detect the transmitted symbols. Besides, we observe that as long as two symbols have the same switching periods, they are highly correlated even if their duty cycles are different.

Figure 8 explains our observation. Two V-PWM symbols have different duty cycles, 75% and 50%. We highlight the matched and the mismatched parts in their pulses $i$ with the green blocks and the red block, respectively. We can see that the total area of the green blocks is larger than that of the red block. This means that the correlation of the $i$-th pulse pair (denoted as $Cor^i_{pulse}$) is positive. In addition, we can check that the correlations of all other pulse pairs are also positive. The correlation of two symbols (denoted as $Cor_{sym}$) can be calculated as the sum of all pulse pairs' correlations:

$$Cor_{sym} = \sum_i Cor^i_{pulse} \tag{3}$$

We can also check that the signs of $Cor^i_{pulse}$ are either all positive or all negative (i.e., their signs are the same). The absolute value of $Cor_{sym}$ will not be canceled, but will accumulate with the increase of pulses.

To validate the above observation, we check the correlations between V-PWM symbols with the same switching periods but with different duty cycles. We generate four V-PWM symbols whose switching periods are the same, and set their duty cycles to 20%, 30%, 40%, and 50%, respectively. For convenience, we denote these symbols as $S_{20}$, $S_{30}$, $S_{40}$, and $S_{50}$. We take the symbol $S_{50}$ as the template symbol and correlate it with the other three symbols ($S_{20}$, $S_{30}$, and $S_{40}$). Figure 9 shows the result, which conforms to our observation. Specifically, we can see that each subfigure in Figure 9 has a sharp peak, which indicates a high correlation. We will further quantitatively evaluate the impact of duty cycle in Section 9.6.

In summary, our design ensures that even if the duty cycle of the symbol used by the receiver is different from that of the symbols actually transmitted, the receiver can still detect these transmitted symbols. In our implementation, the receiver only needs to know the switching periods of V-PWM symbols (i.e., via the appliances' ID), and the duty cycle of the template symbol is fixed to 50% by default.

*6.2.3 Concurrent Transmission.* Although small appliances can now talk to the smart speaker by using their motors, these appliances cannot listen to each other. Therefore, it's infeasible for them to adopt a channel sensing based mechanism (e.g., CSMA) to avoid transmission collision.

MotorBeat supports concurrent transmission. This ability stems from the fact that the V-PWM symbols generated by different random seeds are statistically uncorrelated (or orthogonal) to each other. We can refer to Equation 3 again to understand this. If two V-PWM symbols' switching periods are randomly different, the correlations of their pulse pairs $Cor^i_{pulse}$ will not have the same signs. Therefore, the correlation of these two symbols $Cor_{sym}$ will be canceled with the increase of pulses. Such property allows the receiver to detect symbols from the collided signals separately.

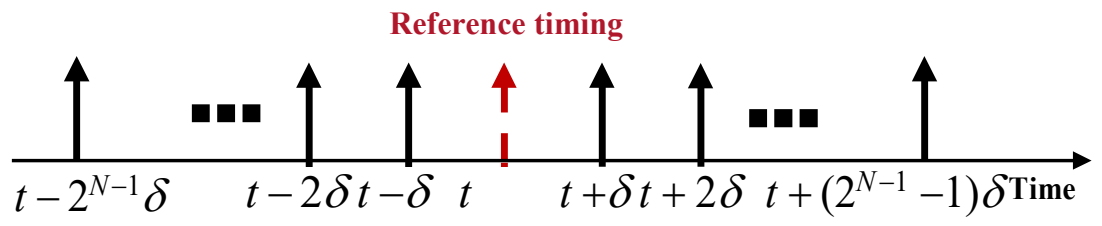

Fig. 11. Encoding data bits by shifting the transmission timing of V-PWM symbols from the reference timing $t$.

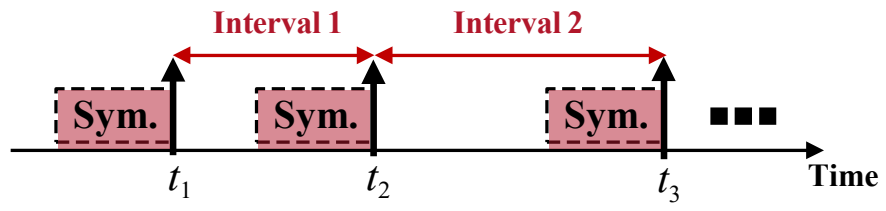

Fig. 12. To relax time synchronization, time intervals between symbols are used for encoding.

In practice, a unique ID is assigned to each appliance to initialize the random state to generate the switching periods $T_{sw}$ of its V-PWM symbol. This design not only ensures the orthogonality of different appliances' symbols, but also can be used to identify appliances because their V-PWM symbols are unique.

## 6.3 Low-SNR Problem Mitigation

Typically, our desired EI signal may be buried by the noise. In order to mitigate this problem, we resort to extending the length (duration) of the transmitted V-PWM symbols, so as to increase the SNR. To detect V-PWM symbols, the MotorBeat receiver takes these symbols as templates and correlates them with the captured acoustic samples. When the correlation is significantly above the noise floor, a V-PWM symbol is assumed to be detected. According to Equation 3, the correlation increases with the number of pulses[17, 57]. When the symbol length increases, the receiver can accumulate more signal energy (i.e., correlation) over pulses. If the transmitted symbol is sufficiently long, the receiver can find a sharp correlation peak, and then detect the presence of the symbol.

We conduct the following experiment to examine the impact of symbol length. A motor is placed 6m away from the microphone and then transmits symbols with different symbol lengths (i.e., 0.1s, 0.5s, and 1s). Figure 10 (a), (b) and (c) display these symbols' correlation results. As we can see, when the symbol length is 0.1s, the correlation peak might be buried by the noise. When the symbol length is increased to 0.5 or 1s, the correlation peaks appear, which means MotorBeat can detect these symbols accurately.

## 6.4 Data Bit Transmission

MotorBeat enables small appliances to transmit data bits to the smart speaker, thus allowing them to report their status, such as remaining battery life and working modes. Inspired by [31, 56], we try to encode data bits into the transmission timing of each V-PWM symbol. Specifically, let us assume that the reference transmission timing of the V-PWM symbol is $t$ (see Figure 11). The transmitter encodes $N$ data bits into a time shift $T_{shift}$, and shifts the transmission timing from the original timing $t$ by $T_{shift}$. The time shift $T_{shift}$ is represented as

$$T_{shift} = K \cdot \delta \tag{4}$$

where $K \in \mathbb{Z}$ whose range is $[-2^{N-1}, 2^{N-1})$, and $\delta$ denotes the time granularity.

However, realizing the above idea requires dedicated time synchronization: the transmitter and the receiver share the same clock and make an appointment on the reference timing $t$ . Apparently, time synchronization is almost infeasible for most low-end appliances who are isolated from the Internet.

To relax the requirement for time synchronization, we exploit the relative interval between two adjacent symbols to encode data bits instead. In this case, the transmission timing of the current symbol can be viewed as the reference timing of the next symbol. Figure 12 illustrates our idea. The transmitter deliberately sends multiple same symbols after different time intervals to encode data bits. The intervals $T_{interval}$ are represented as

$$T_{interval} = T_{min} + T_{shift} \tag{5}$$

where $T_{min}$ denotes the minimum symbol interval. To avoid the overlap between symbols, the minimum symbol interval $T_{min}$ is set to $\frac{5}{4}L_{sym} + 2^{N-1} \cdot \delta$ in our design, where $L_{sym}$ denotes the length of V-PWM symbol.

Next, we analyze the data rate $R_b$ MotorBeat can achieve. We assume $M$ successive symbols are sent during each round of transmission. At first glance, it seems that $M$ successive symbols have $M - 1$ intervals and thus $(M - 1) \cdot N$ bits will be conveyed. However, in addition to symbol intervals, each V-PWM symbol itself also implicitly contains the appliance's ID information, based on which we can identify the transmitter of data bits. In our design, a ground of 16 bits is used to represent an appliance's ID[7]. Then, the total number of bits transmitted is $(M - 1) \cdot N + M \times 16$ bits. The average duration of each round of transmission can be calculated as $L_{sym} + (M - 1) \cdot \mathbf{E}[T_{interval}]$, where $\mathbf{E}[T_{interval}]$ denotes the expectation of $T_{interval}$, roughly equaling $T_{min}$. Finally, the average data rate can be given by

$$R_b = \frac{(M - 1) \cdot N + M \times 16 \text{ bits}}{L_{sym} + (M - 1) \cdot \mathbf{E}[T_{interval}]} \tag{6}$$

As we can see, $\mathbf{E}[T_{interval}]$ increases with $N$ exponentially. With the increase of $N$, $R_b$ is gradually dominated by $N$, and decreases exponentially. This indicates that we should not increase $N$ greedily, but the dedicated optimization of data rate should be conducted.

In practice due to clock skew error and multipath effect, time granularity can not be arbitrarily small. To compromise these problems, we resort to setting a low time granularity, i.e., $\delta = 20$ ms. Meanwhile, symbol length $L_{sym}$ is 1 second to ensure a high detection accuracy, and $M$ is set to 4. Given those settings, $N$ should be 3, yielding an optimal data rate of 14.6 bps. Note that this data rate is calculated under a conservative setting. As we will see in Evaluation, the data rate can also be up to hundreds bps under certain settings.

## 7 RECEIVER DESIGN

The receiver, i.e., smart speaker, gets acoustic samples from its microphone, and then detects V-PWM symbols in the following ways: The receiver first normalizes previously-known V-PWM symbols to $[-1, 1]$. Next, it takes each normalized symbol as a template, and correlates it with the acoustic samples. The receiver then measures the mean ($\mu$) and the standard deviation ($\sigma$) of the noise of the absolute correlation, and sets a detection threshold as five standard deviations above the mean (i.e., $\mu + 5\sigma$). If the absolute correlation is larger than the threshold, this V-PWM symbol is assumed detected. After that, the receiver can output two types of information as follows:

- **Heartbeat.** As mentioned before, each appliance's ID is used as the random seed of its V-PWM symbol. If one V-PWM symbol is detected, we then know the corresponding appliance is ON at present. In other words, the receiver can output a three-tuple of information: **<Appliance ID, timestamp, ON>**. We view this kind of information as the heartbeats of appliances.
- **Data Bit.** In our design, one appliance may transmit data bits by sending multiple successive symbols whose intervals are encoded with messages. That is to say, the receiver may output multiple heartbeats of one appliance sequentially. To receive the messages, the receiver only needs to extract the time intervals from the timestamps of one appliance's heartbeats, and decode data bits from these intervals. Then, the receiver can also output the messages in a three-tuple form: **<Appliance ID, timestamp, Data Bits>**.

It is worth noticing that correlation is a costly operation. When there are many appliances, the smart speaker as the receiver needs to separately correlate each appliance's V-PWM symbol with the received signal. The corresponding computational cost may be too much for the receiver to afford. In order to mitigate this problem, we have the following optional enhancements to the receiver design: (1) For heartbeat reception, one may let appliances broadcast heartbeats at a large time interval (e.g., 10 s). This time interval is also known to the receiver.

[7]Of course, we can use more bits to represent an appliance's ID to generate more IDs. However, finite coding space can only accommodate limited V-PWM symbols that are orthogonal to each other. The coding space is finite because of finite V-PWM symbol length. If we increase the number of IDs regardless, their corresponding V-PWM symbols will not maintain sufficient Hamming distances. This means that the orthogonality among V-PWM symbols will not hold any more. According to our experience, 16 bits provides a reasonable number of IDs ($2^{16}$) and holds a good orthogonality among symbols.

After detecting one appliance, it is safe for the receiver to skip detecting this appliance within this time interval. By doing so, we can avoid brute-force detection and thus reduce the correlation operations. (2) For data bit reception, we may define a special V-PWM symbol as the preamble for all appliances. Any appliance that prepares to communicate with the receiver should send this preamble before sending its own data. When the receiver starts receiving data bits, it will first correlate the signal with the preamble rather than all the possible V-PWM symbols. In this way, the receiver can be notified about the existence of communication from an appliance. After that, the receiver continues to correlate all possible symbols with the subsequent acoustic signals to decode data.

## 8 PRACTICAL ISSUES

### 8.1 Configuration of Smart Speaker

MotorBeat requires users to associate their small appliances with the smart speaker manually. Specifically, we envision a typical scenario as follows: Manufacturers print a QR code in the appearance of small appliances. Each QR code contains the necessary configuration information, such as ID (the V-PWM symbol's random seed), the parameters of the modulation and the encoding. To associate a new appliance with the smart speaker, the user takes the phone to scan the QR code and synchronize the configuration to the smart speaker via WiFi.

### 8.2 Multipath Effect

Here, we discuss the effect of multipath on MotorBeat. In Section 9.9, we will further evaluate it quantitatively.

For the heartbeat detection, the impact of the multipath effect is negligible. When the correlation is larger than a threshold, we assert a heartbeat is detected. In other words, we only care about the energy level of the signals. As long as the energy of one path is strong enough, we can detect heartbeats.

For the data bit decoding, multipath has a negative effect. The data bits are embedded into the time interval between two symbols. It is possible that the receiver detects path A in the first symbol, but detects path B in the second symbol. In this case, the detected time interval is polluted by the time-difference-of-arrival (TDOA) between path A and B. MotorBeat adopts a relatively large time granularity of time interval ($\delta$ = 20 ms) to alleviate this problem. The intuition is that the large time interval provides a wide guard interval that can tolerate the TDOA noise introduced by multipath propagation.

### 8.3 Hardware Modification

To apply MotorBeat, appliances with different hardware require hardware modification at different extents.

- For appliances equipped with MCUs, the modification is merely to rewire the DC motor driver to GPIO pins of MCU (similar to our implementation in Section 9.1). One may concern that generating V-PWM symbols via MCU may require a high clock frequency and then introduce extra power consumption. Actually, MCUs running at low clock frequencies are sufficient to generate our V-PWM symbols: The change of GPIO pins' status (ON or OFF) is triggered by the MCU's timer. In other words, each pin acts as a Digital-to-Analog Converter (DAC) whose sampling rate is determined by the clock frequency [5]. In our design, the switching frequencies of V-PWM symbols are between 500 Hz and 2000 Hz and the main bandwidth of V-PWM symbols is typically less than 12 kHz. Therefore, it is easy for a MCU running at a low clock frequency (e.g., 16 MHz) to achieve the required Nyquist sampling rate (e.g., 24 kHz).
- For appliances without MCUs but with PWM drivers, we believe our work would motivate chip manufacturers to modify the firmware of PWM drivers to support V-PWM. If so, appliances equipped with such drivers can keep broadcasting their heartbeats during operation. However, these appliances can not transmit data bits because they have no MCU to control time intervals between V-PWM symbols.

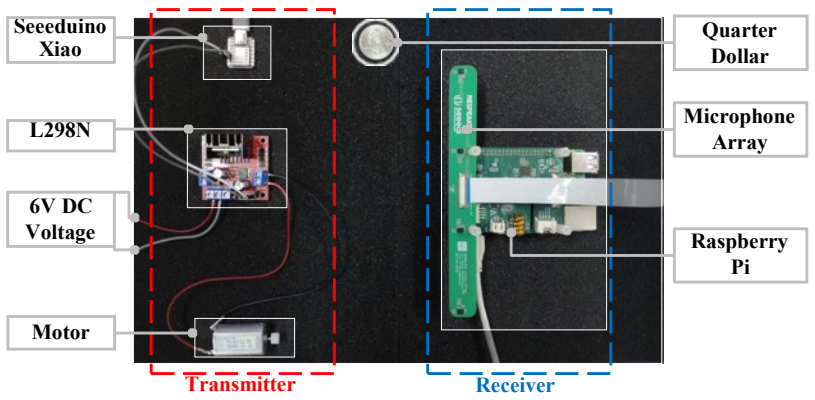

Fig. 13. The prototype of MotorBeat.

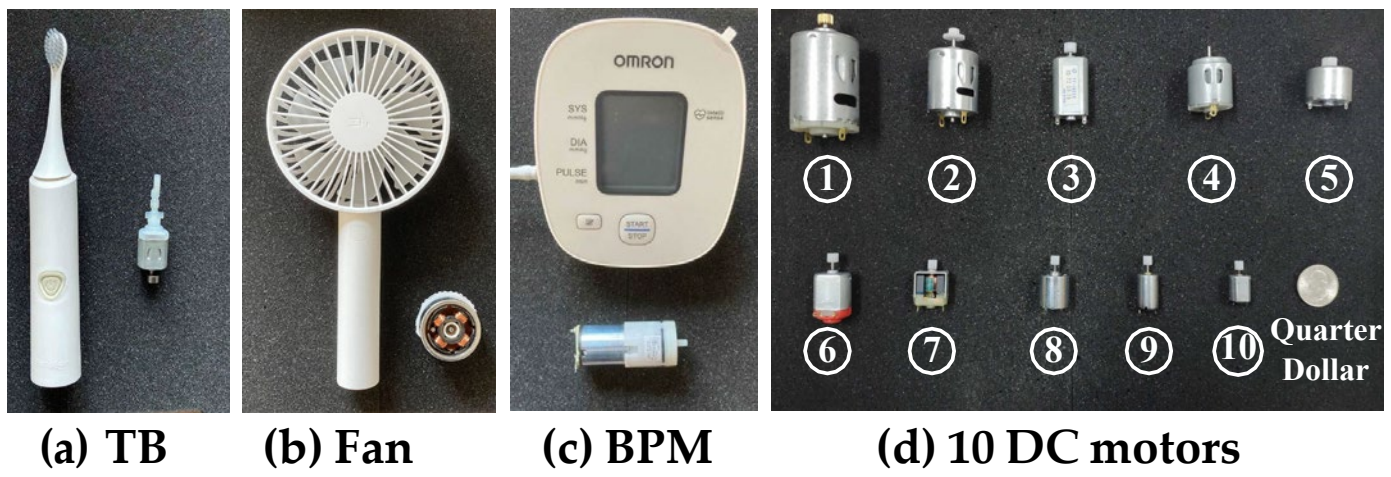

Fig. 14. Three appliances and ten DC motors.

- For appliances that directly use analog voltage to drive their motors, extra hardware installation is required, such as MCU or dedicated V-PWM driver.

.

### 8.4 Brushless Motor

MotorBeat can be applied to both brushed motors and brushless motors. In a brushless motor, the coils (stator) are fixed while the permanent magnets (rotor) can be rotated. Different from brushed motors which use brushes to automatically switch the polarity of electromagnet, brushless motors typically adopt dedicated drivers which contain a sensor (e.g., hall effect or back-EMF sensor) to measure the rotor position and then to choose a proper polarity of electromagnet. In spite of these differences, the underlying principles of brushed and brushless motors are the same: using dynamic voltage to generate dynamic magnetic fields and then to attract or repel the magnets. In other words, brushless motors also produce the EI signal which can be modulated by the voltage (see Section 4). To drive brushless motors to transmit V-PWM symbols, we simply modulate the input DC voltage to the brushless motor driver (e.g., three-phase inverter) according to V-PWM.

## 9 EVALUATION

### 9.1 Implementation

**Hardware.** Figure 13 shows the prototype of MotorBeat. Since we have no access to small appliances' embedded software, we then directly drive their motors by connecting them to our transmitter prototype. We use a Seeeduino Xiao ($4.9) [59] and an L298N module ($1.6) to drive the motors. Specifically, the motors are powered by the L298N module. The voltage is 6V. The Seeeduino Xiao has two GPIO pins connected to the L298N module. The Seeeduino can switch the motors on and off by toggling the state of the pins. The V-PWM symbol's switching periods are uniformly distributed between 0.5 ms and 2 ms, smaller than the time constant of DC motors. The corresponding switching frequencies are between 500 Hz and 2000 Hz. To modulate the motor's sound, the Seeeduino toggles the pins according to its V-PWM symbol. For the receiver, we use a ReSpeaker 4-mic linear array [58] to act as a smart speaker to receive the acoustic signal with sampling rate 24 kHz. The array is sitting on top of a Raspberry Pi 4 Model B, on which we run the software to detect and decode the symbols.
**Software.** The signal processing is written in Python. We use the sliding-window method to process the acoustic signal stream chunk by chunk. To ensure that a sliding window can cover the whole V-PWM symbol, we set the window length to $1.25L_{sym}$. The sliding windows are overlapped because the symbol boundary is unknown. The sliding size is $0.25L_{sym}$. To detect all V-PWM symbols, we correlate each symbol with each window of samples.

### 9.2 Experimental Methodology

Figure 14(a) shows three appliances used for evaluation: LangTian LT-G8 electric toothbrush ($10.4), ZMI AF215 mini fan ($9.1), and OMRON U10L blood pressure monitor ($30.6). For convenience, these three appliances are

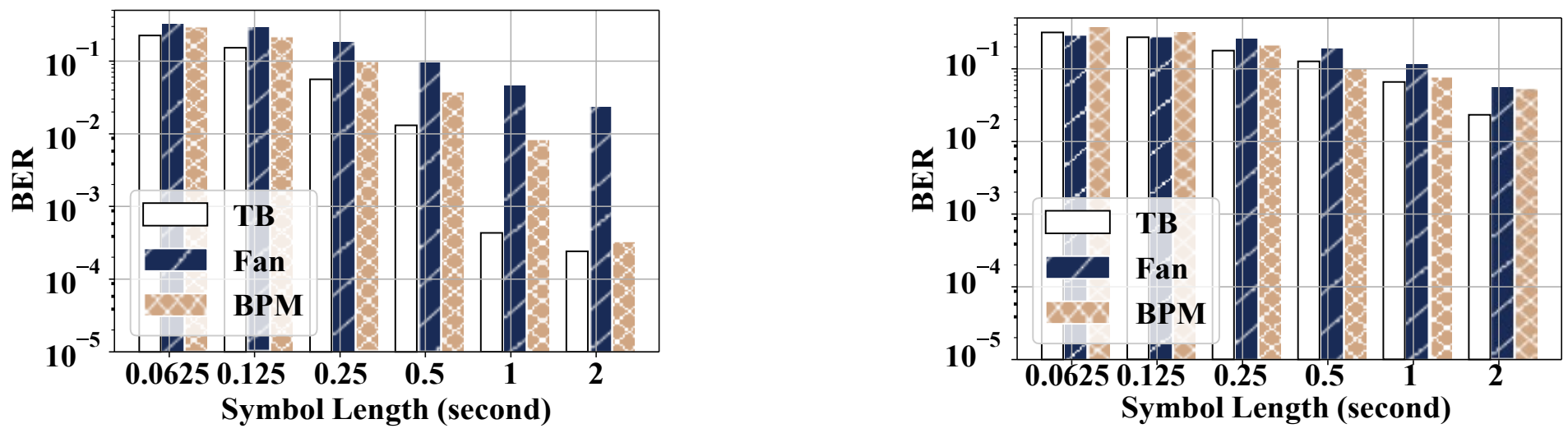

(a) Distance = 2 m

(b) Distance = 10 m

Fig. 15. BER vs. symbol length (time resolution = 20 ms).

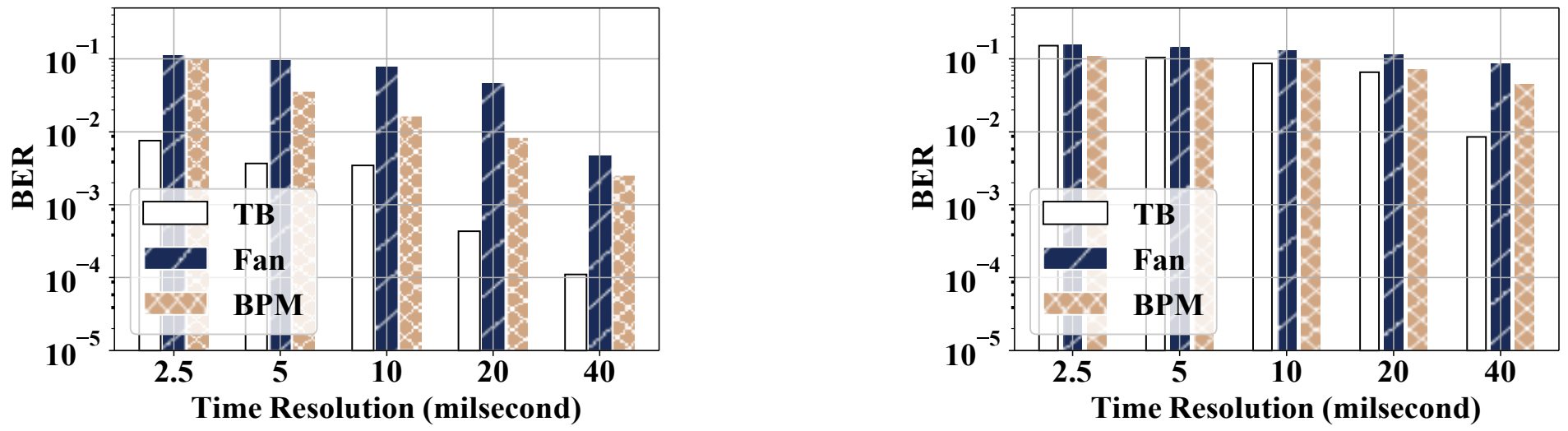

(a) Distance = 2 m

(b) Distance = 10 m

Fig. 16. BER vs. time resolution (symbol length = 1 s).

abbreviated as TB, Fan, and BPM, respectively. Their motors are also displayed. Note that the motors of TB and BPM are brushed while the motor of Fan is brushless. During evaluation, these appliances' motors are kept in their home (original) positions and these appliances are placed facing the receiver at different distances. To further show that MotorBeat can be widely applied to DC motors, we use another ten different motors for evaluation, shown in Figure 14(d). The average price of motors is $0.39. During evaluation, each volunteer holds one motor, while facing the receiver. We conduct experiments in a classroom. The background noise is around 40-48 dB.

In the following, we first evaluate the overall performance of transmitting data bits (Section 9.3). Because data bits are encoded into time intervals between V-PWM symbols (i.e., heartbeats), reliable transmission performance highly depends on accurate heartbeat detection. Section 9.4 then presents heartbeat detection accuracy. Next, we use ten motors to evaluate MotorBeat (Section 9.5). To present **3C** features, we conduct head-to-head experiments and deliver the results of each goal (Section 9.6). Furthermore, we evaluate the computing cost of the receiver on the Raspberry Pi (Section 9.7). We also evaluate MotorBeat in the mobile scenario (Section 9.8). Finally, the performance in different multipath cases are evaluated (Section 9.9).

## 9.3 Overall Performance

This subsection evaluates the performance of transmitting data bits. Three appliances, TB, Fan, and BPM, are used for evaluation. We place the receiver 2m and 10m away from the transmitter, and then separately study how time resolution and symbol length affect Bit Error Rate (BER). Since our focus here is the communication performance, we use V-PWM symbols to transmit raw bits and evaluate the BER.

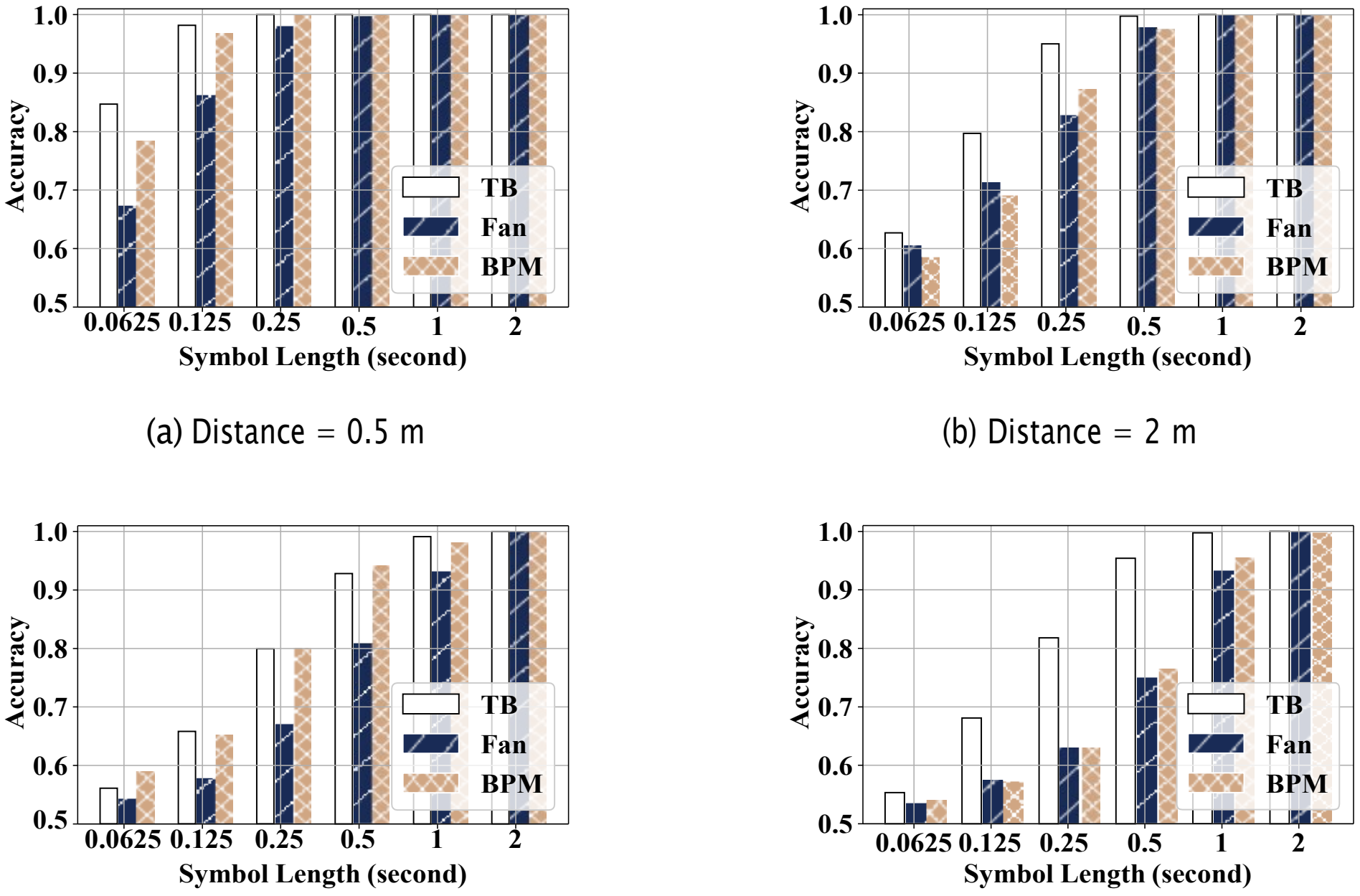

(a) Distance = 0.5 m
(b) Distance = 2 m
(c) Distance = 6 m
(d) Distance = 10 m

Fig. 17. Heartbeat detection accuracy vs. symbol length in different distance settings.

Figure 15 shows the result with respect to symbol length. With the increase of symbol length, the correlation peaks tend to be sharper. Therefore, the receiver can more precisely determine the arrival time of each symbol. We vary the symbol length from 0.0625 s to 2 s to measure BER. Meanwhile, the time resolution is fixed to 20 ms. The throughputs under these settings are 195.8 bps, 102.6 bps, 53.7 bps, 28.1 bps, 14.6 bps, and 7.6 bps, respectively. (The throughputs are calculated by Equation 6.4. $M$ is set to 4 by default and $N$ is set to a value that maximizes the throughput.) We observe that the BER decreases as the symbol length increases. When the symbol length is less than 0.5s, the BERs of some appliances are higher than 10%, which might not be desirable. The problem can be alleviated by increasing the symbol length. When the symbol length is 2s, these BERs are not more than 5.4%. As an example, the BER of the TB at 2m and 10m is 0.024% and 2.3%, respectively.

Figure 16 shows the result with respect to time resolution $\delta$. As discussed in Section 8.2, a larger time resolution provides a wider guard interval and mitigates the timing error and multipath effect. We then vary the time resolution from 2.5ms to 40ms to evaluate BER. The symbol length is fixed to 1s. The throughputs under these settings are 16.4 bps, 15.8 bps, 15.2 bps, 14.6 bps, and 14.0 bps, respectively. As expected, the BER decreases with the increase of the time resolution. When the time resolution is 40 ms, the payload of the TB can be accurately decoded. The BERs are only 0.011% at 2m, and 0.85% at 10m.

Apparently, the performance also varies across appliances. The main reasons are two-fold: (1) As we can see in Figure 14(a)-(c), the motors of these appliances are quite different. Naturally, the sounds generated by these motors and their signal quality are also different. (2) These appliances attach different gearboxes to their motors for different purposes, such as the toothbrush head for the TB, the blades for the Fan, and the air pump for the BPM. These gearboxes introduce varying degrees of noise. For example, the Fan's blades will generate significant aerodynamic noise. This explains why the Fan performs worst among these appliances.

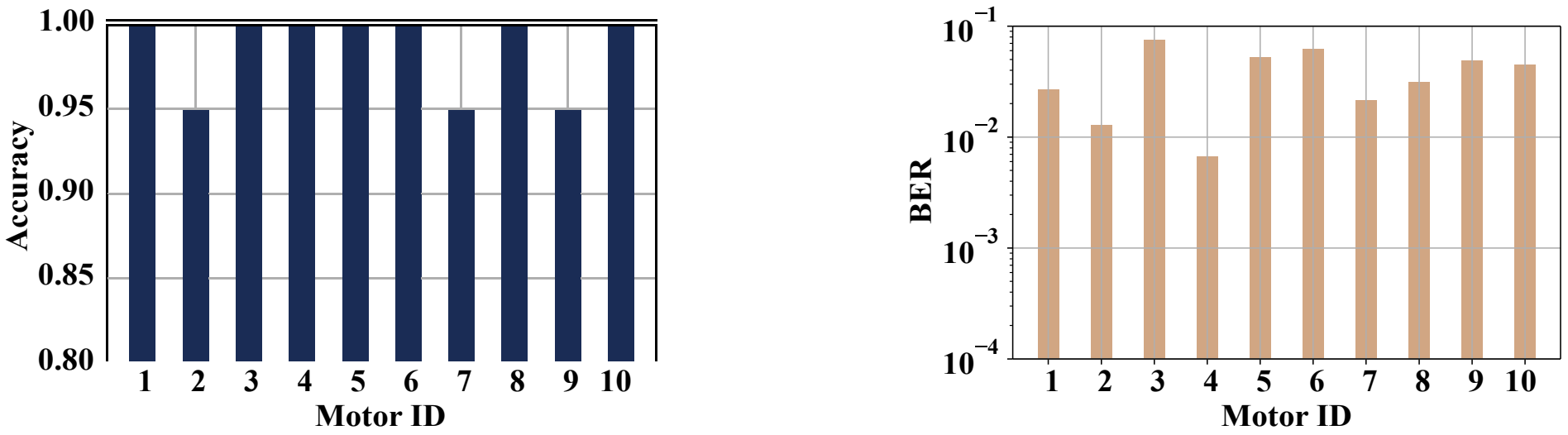

(a) Heartbeat accuracy

(b) BER

Fig. 18. Performance across 10 motors (Distance = 6 m).

## 9.4 Heartbeat Detection Accuracy

This subsection studies the impact of the distance and the symbol length on the detecting heartbeats (i.e., detecting V-PWM symbols). We place the receiver 0.5m, 2m, 6m, and 10m away from the transmitter. For each distance setting, we further vary the symbol length from 0.0625s to 2s to measure the accuracy of detecting heartbeats. Figure 17 shows the results.

By comparing Figure 17(a)-(d), we can find that the accuracy decreases as the distance increases. This is expected since the longer distance means more serious signal attenuation. To combat the problem of attenuation, we can increase the symbol length to extend the communication range. For example, when the symbol length increases to 2s, the detection performance is nearly perfect, and no error occurs in all distance settings. On the other hand, when the distance is no more than 2m, we can use symbols with shorter lengths, such as 0.5s or 1s, which also maintain a reasonable accuracy ($\geq$0.973). Again, because different appliances generate different levels of noise, their detection performance is also different.

## 9.5 Performance across 10 DC Motors

Here, we want to show the feasibility of applying MotorBeat to various DC motors. We evaluate MotorBeat's performance on 10 different motors shown in Figure 14(d). We place each motor 6m away from the receiver. Figure 18(a) shows the result of the heartbeat detection accuracy. In this experiment, the symbol length is fixed to 1s. The heartbeats of these motors can be accurately detected. The worst accuracy (Motor 7) can still reach 0.924. Figure 18(b) shows the result of the BER. In this experiment, the symbol length and the time resolution are fixed to 1s and 20ms, respectively. All the BERs are less than 10%, and the average BER across all motors is only 3.8%.

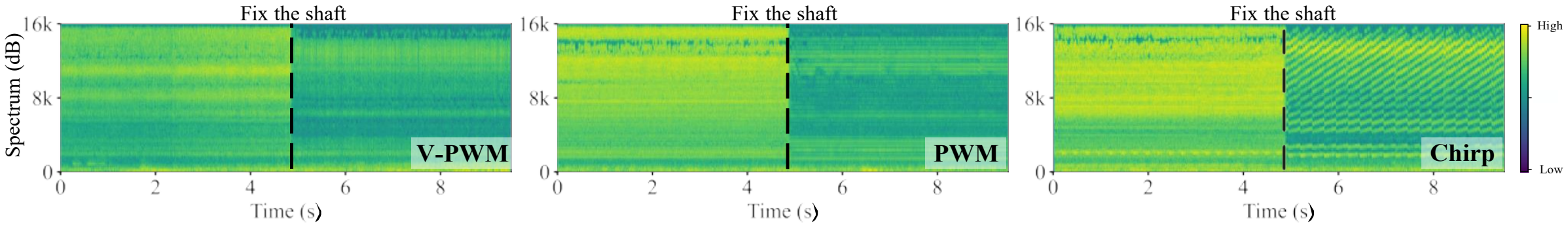

(a) (b) (c)

Fig. 19. Spectrums of motor sounds modulated by (a) V-PWM, (b) PWM and (c) Chirp.

## 9.6 3C Goals

We obtained the IRB approval to conduct the following experiments.

Table 2. The average accepted scores of the sounds generated by three modulation methods.

| Modulation method | Motor A | Motor B | Motor C |
|---|---|---|---|
| V-PWM (MotorBeat) | **4.2** | 3.8 | **4.5** |
| PWM (Standard) | 4.1 | **4.0** | **4.5** |
| Chirp (Bleep) | 3.0 | 2.2 | 2.8 |

*9.6.1 Comfortable.* Here, we evaluate the acceptance of the sounds produced by MotorBeat. We compare MotorBeat with two other methods: standard PWM adopted by most appliances, and chirp-like modulation adopted by Bleep [8]. Figure 19 shows the spectrums of motor's sound modulated by these methods, respectively. The running motor will produce both EI signal and MI signal. To clearly display the EI signal, we manually fix the shaft (rotor) to avoid MI signal after the motor runs for a while. As we can see, the sound modulated by Chirp generates harmonics whose frequencies increase gradually. This sound is too conspicuous to neglect. On the other hand, MotorBeat distributes the energy of harmonics over a wide frequency range and thus the sound modulated by V-PWM is comfortable to hear.

We further invite 10 volunteers to join this experiment. We place each motor 0.5m away from volunteers, and drive the motor with three modulation methods separately. After hearing each sound, each volunteer is then asked to rate the acceptance: 5=acceptable, 4=slightly acceptable, 3=mild, 2=slightly unacceptable, and 1=unacceptable. Table 2 shows the average accepted scores of these sounds. The results show that Bleep seriously disturbs the users, and has the lowest acceptance among these three methods. Meanwhile, MotorBeat introduces no additional annoying sounds, and almost has the same acceptance as the standard PWM.

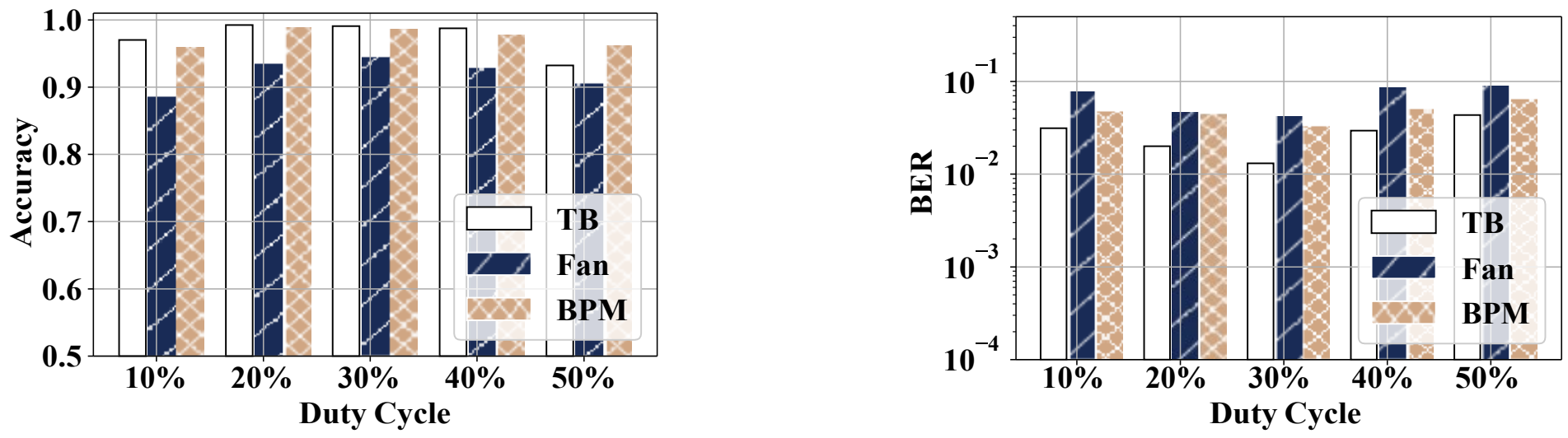

(a) Heartbeat accuracy

(b) BER

Fig. 20. Performance with varying duty cycles.

*9.6.2 Compatible.* The user may change the working mode of the appliances at will. The duty cycle of transmitted V-PWM symbols is unknown to the receiver. This experiment intends to show that even if the duty cycle of the symbol used by the receiver is different from that of the symbols actually transmitted, the receiver can still detect these transmitted symbols. Each appliance is 6m away from the receiver, and repeatedly transmits its V-PWM symbols (length = 1s) with a fixed duty cycle of 30%. The receiver detects these V-PWM symbols with varying duty cycles: 10%, 20%, 30%, 40%, and 50%. Figure 20 (a) and (b) show the results. As expected, the receiver achieves the best accuracy when the duty cycle is 30%. This is because the symbol with duty cycle 30% perfectly matches the transmitted symbol. When the receiver uses symbols with other duty cycles, the performance only suffers a slight degradation.

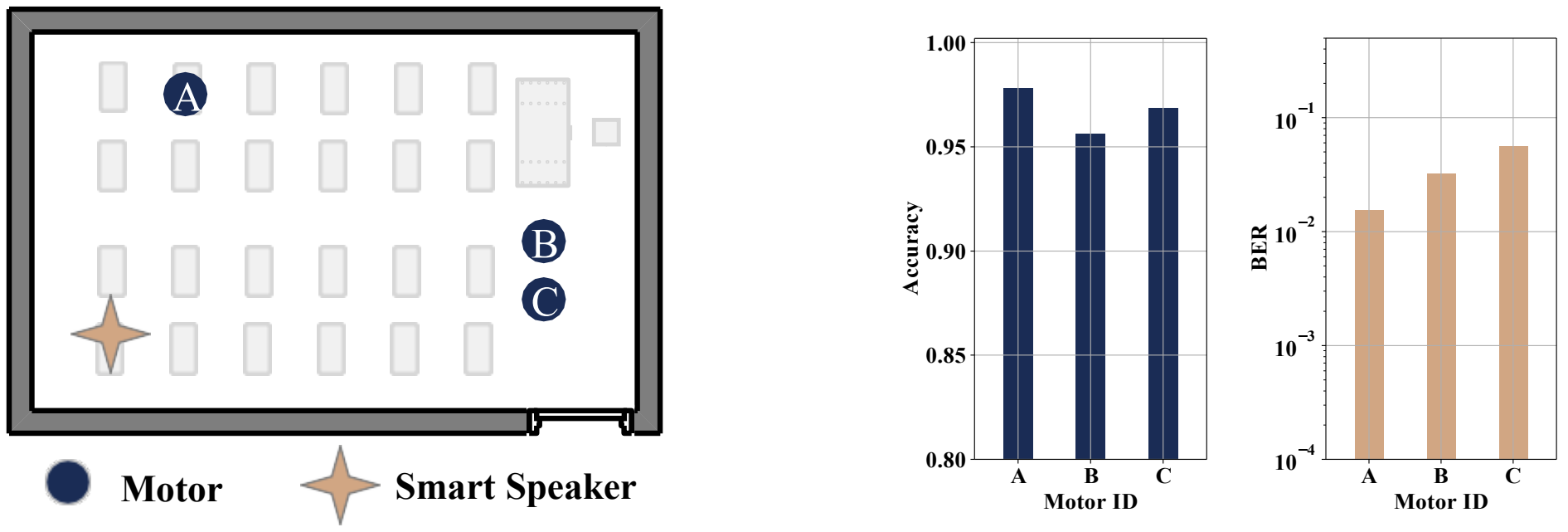

(a) Deployment (8m x 5m) (b) Heartbeat (c) BER

Fig. 21. Performance when 3 motors transmit concurrently.

*9.6.3 Concurrent.* Here, we evaluate the performance in the multi-source scenario. We place three motors (Motor 3, 6, and 8 in Figure 14) in a room with a size of 8m x 5m, as shown in Figure 21(a), and let them transmit simultaneously. The other settings are the same as those in Section 9.5. Figure 21 (b) and (c) show the results. In general, MotorBeat achieves a satisfactory performance. All heartbeat detection accuracies are above 0.95, and the worst BER is still less than 5.6%.

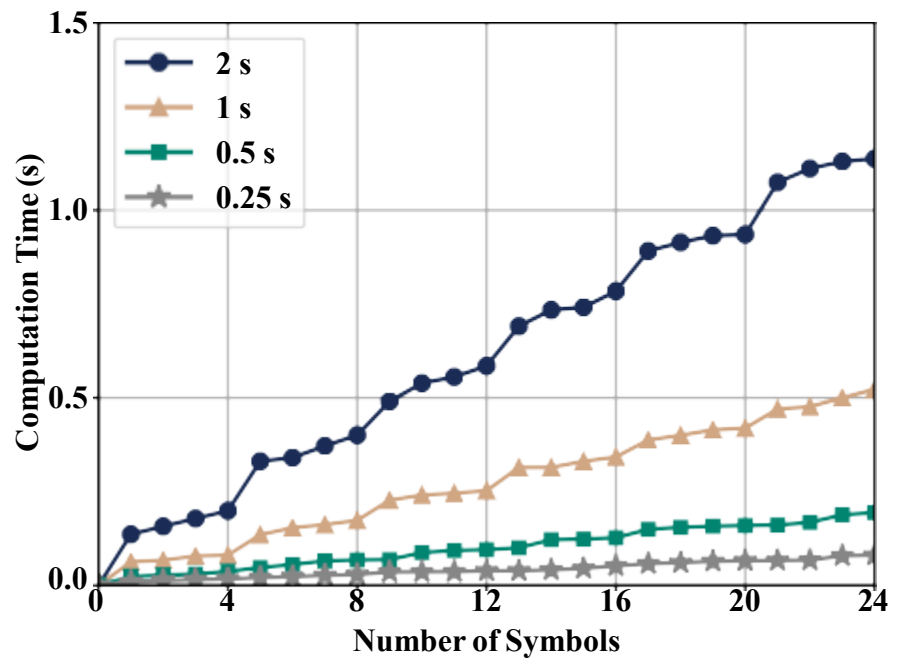

Fig. 22. Computing time cost.

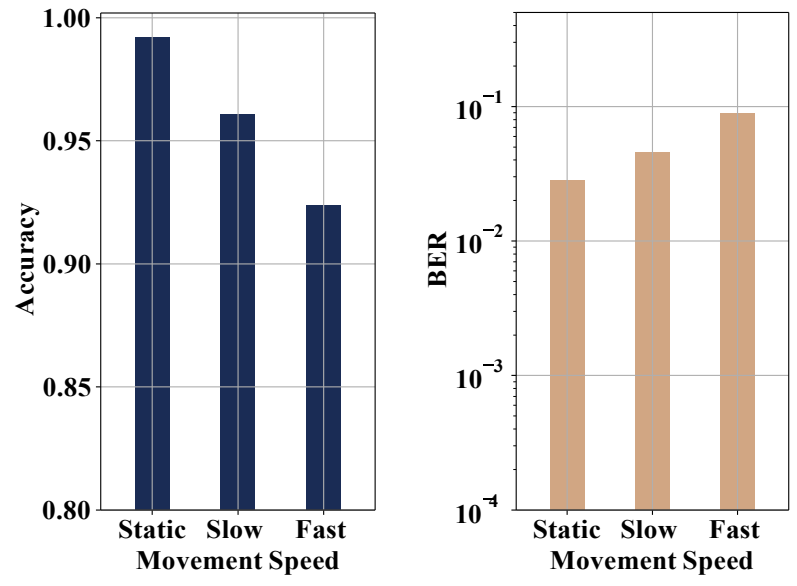

(a) Heartbeat (b) BER

Fig. 23. Performance vs. speed.

## 9.7 Computational Efficiency

Here, we evaluate the computation efficiency of MotorBeat. We run the MotorBeat receiver on a Raspberry Pi 4B whose CPU has four cores. We then use four processes to compute correlations simultaneously. Figure 22 shows the computational time as a function of the number of symbols. We use a sliding window method to process the acoustic samples. In order to process the acoustic signal in real time, MotorBeat should detect all V-PWM symbols within the sliding size (0.25 $L_{sym}$). For example, for the 2s symbols, MotorBeat needs to detect them all within 0.5s. By referring to Figure 22, we find that MotorBeat takes 0.49s to detect 9 symbols. In other words, MotorBeat can detect 9 symbols with a length of 2s in real time.

## 9.8 Impact of Mobility

Some small appliances, such as toothbrushes, might not be static during usage, so this section studies how the movement of the transmitter affects MotorBeat. A volunteer is asked to stand 6m away from the receiver, and hold and move the toothbrush with three speeds: static, slow (about 0-0.2m/s), and fast (about 0.5-1m/s). In this experiment, the symbol length is 1s, and the time resolution is 20 ms. Figure 23 shows that the performance decreases with the increase of speed. This is because the movement introduces an additional Doppler effect, which undermines the correlation between the symbol and the acoustic samples. The faster the transmitter moves, the more serious the Doppler effect will be. From our measurements, when the speed increases to 2 m/s, the receiver will fail to detect the symbols. Nevertheless, the typical speed of small appliances is no more than 1 m/s at home.

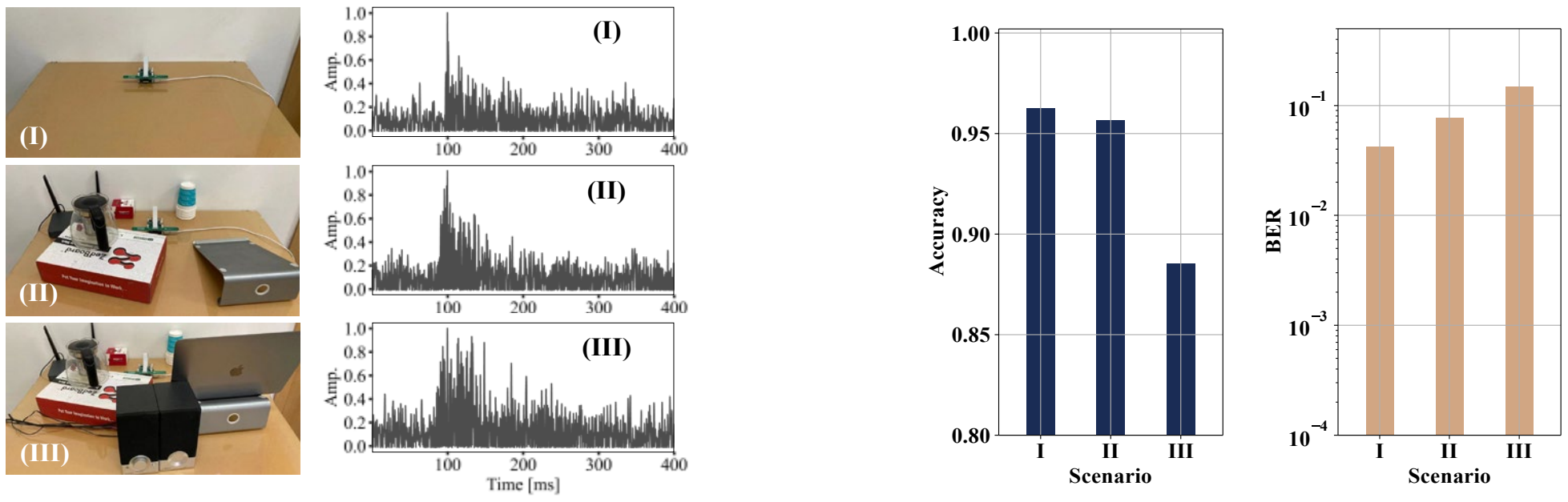

(a) Three Scenarios (b) Heartbeat (c) BER

Fig. 24. Performance vs. multipath severities

## 9.9 Impact of Multipath

Here, we evaluate how multipath affects MotorBeat. We place the receiver in three multipath scenarios I, II, and III. Figure 24(a) displays these scenarios as well as their corresponding Channel Impulse Responses (CIRs)[8]. Apparently, Scenario I's channel is the cleanest, while in Scenario III, the line-of-sight path is completely blocked and its channel experiences the most severe multipath effect. In this experiment, the motor is 6m away from the receiver, the symbol length is 1s and the time resolution is 20 ms. Figure 24 (b) and (c) show the performances in terms of heartbeat detection and data bit decoding, respectively. As expected, the performance degrades with the severity of multipath. However, the two applications have different sensitivities to the multipath effect. By comparing the results of Scenarios I and II, we can find that the heartbeat detection accuracy only suffers a slight degradation, from 0.963 to 0.957, while the BER experiences a relatively considerable increase, from 4.2% to 7.6%. In the NLOS case (Scenario III), the performance degrades with the accuracy of 0.885 and the BER of 14.7%.

# 10 LIMITATIONS AND FUTURE WORK

**One-Way Communication.** MotorBeat enables small appliances only to speak, not to listen. Therefore, the receiver cannot reply ACKs to the small appliances. To improve the communication reliability, one may adopt an error-correction code (ECC), such as hamming code or cyclic redundancy check. Another option is to increase the number of transmissions of a same packet. Apparently, both solutions trade-off data rate for reliability. Given the low date rate of MotorBeat, we need to carefully set the code rate of ECC or the re-transmission times.

[8]Here, CIR is obtained from the correlation between the symbol and the acoustic samples.

**Through-Wall Communication.** The acoustic signal is dramatically attenuated if propagated through a wall. Given the fact that the SNR is intrinsically low, it is difficult to connect an appliance to a smart speaker when they are in different rooms or floors.

**Microphone Array and Beamforming.** Most smart speakers contain multiple microphones. We only use one microphone. It is possible to further improve our performance by using beamforming techniques.

**Localization and Motion Sensing.** Recent works [1, 13, 61, 69] demonstrate the feasibility of localizing sounds with a single smart speaker. In our case, small appliances are also acoustic sources, and thus can be localized. What's more, the receiver can measure the CIR of each small appliance from the correlation function between the previously-known symbol and the audio samples, such as Figure 24(a). Therefore, a wide range of wireless sensing techniques [27, 66–68, 70, 74, 76] might be adopted to sense small appliances' motions.

## 11 CONCLUSION

A critical problem that impedes the development of smart home is that most home appliances are cut off from the Internet. To fill this gap, we propose MotorBeat, a novel motor-based communication approach. MotorBeat exploits DC motors, which widely exist in small appliances, to connect small appliances to the Internet. We hope that MotorBeat can pave the way to the vision of smart home, and open up a wide range of applications.

## ACKNOWLEDGMENTS

We thank the anonymous reviewers for their insightful comments. This research is supported by National Natural Science Fund of China No. U21B2007, and Tsinghua University - Meituan Joint Institute for Digital Life.